\documentclass[authoryear]{elsarticle}
\usepackage{microtype}

\usepackage{tcolorbox}[boxsep=1mm]

\usepackage{graphicx}
\usepackage{booktabs}
\usepackage{hyperref}

\makeatletter
\def\MT@warn@unknown{}
\makeatother %

\begin{document}

\title{Architecting Complex, Long-Lived Scientific Software }
\author[1]{Neil A. Ernst}\corref{cor1}
\ead{nernst@uvic.ca}
 \cortext[cor1]{Corresponding author}
 
\author[2]{John Klein}
\ead{jklein@sei.cmu.edu}

\author[3]{Marco Bartolini}
\ead{Marco.Bartolini@skao.int}

\author[4]{Jeremy Coles}
\ead{j.coles@mrao.cam.ac.uk}

\author[3]{Nick Rees}
\ead{Nick.Rees@skao.int}

\address[1]{University of Victoria, Canada}

\address[2]{Carnegie Mellon University Software Engineering Institute, USA }
 
\address[3]{SKA Observatory, United Kingdom}

\address[4]{University of Cambridge, United Kingdom}

\begin{abstract}
Software is a critical aspect of large-scale science, providing essential capabilities for making scientific discoveries.
Large-scale scientific projects are vast in scope, with lifespans measured in decades and costs exceeding hundreds of millions of dollars. 
Successfully designing software that can exist for that span of time, at that scale, is challenging for even the most capable software companies. 
Yet scientific endeavors face challenges with funding, staffing, and operate in complex, poorly understood software settings. 
In this paper we discuss the practice of early-phase softwarearchitecture in the Square Kilometre Array Observatory's Science Data Processor. The Science Data Processor is a critical software component in this next-generation radio astronomy instrument. 
We customized an existing set of processes for software architecture analysis and design to this project's unique circumstances.
We report on the series of comprehensive software architecture plans that were the result.
The plans were used to obtain construction approval in a critical design review with outside stakeholders.
We conclude with implications for other long-lived software architectures in the scientific domain, including potential risks and mitigations. 

\end{abstract}

\maketitle

\section{Introduction}
The Square Kilometre Array (SKA) Project is developing two next-generation radio telescopes which, when completed, will be the world's largest radio telescopes and have unprecedented sensitivity. To build and operate the telescopes, seven countries ratified a treaty to form an intergovernmental organization, the Square Kilometre Array Observatory (SKAO) (with nine more countries showing an interest to join).
The SKAO's telescopes will be able to answer many fundamental questions including describing the formation of the first stars in the universe and the nature of gravity \citep{Labate2019}.

A critical part of the SKAO will be its software systems, which will control the physical instruments that collect the data, and include an exascale computing system to process and analyze the results. It is therefore critical to carefully plan, design, implement, and operate SKAO's software systems.

\begin{figure*}[bht]
    \centering
    \includegraphics[width=\textwidth]{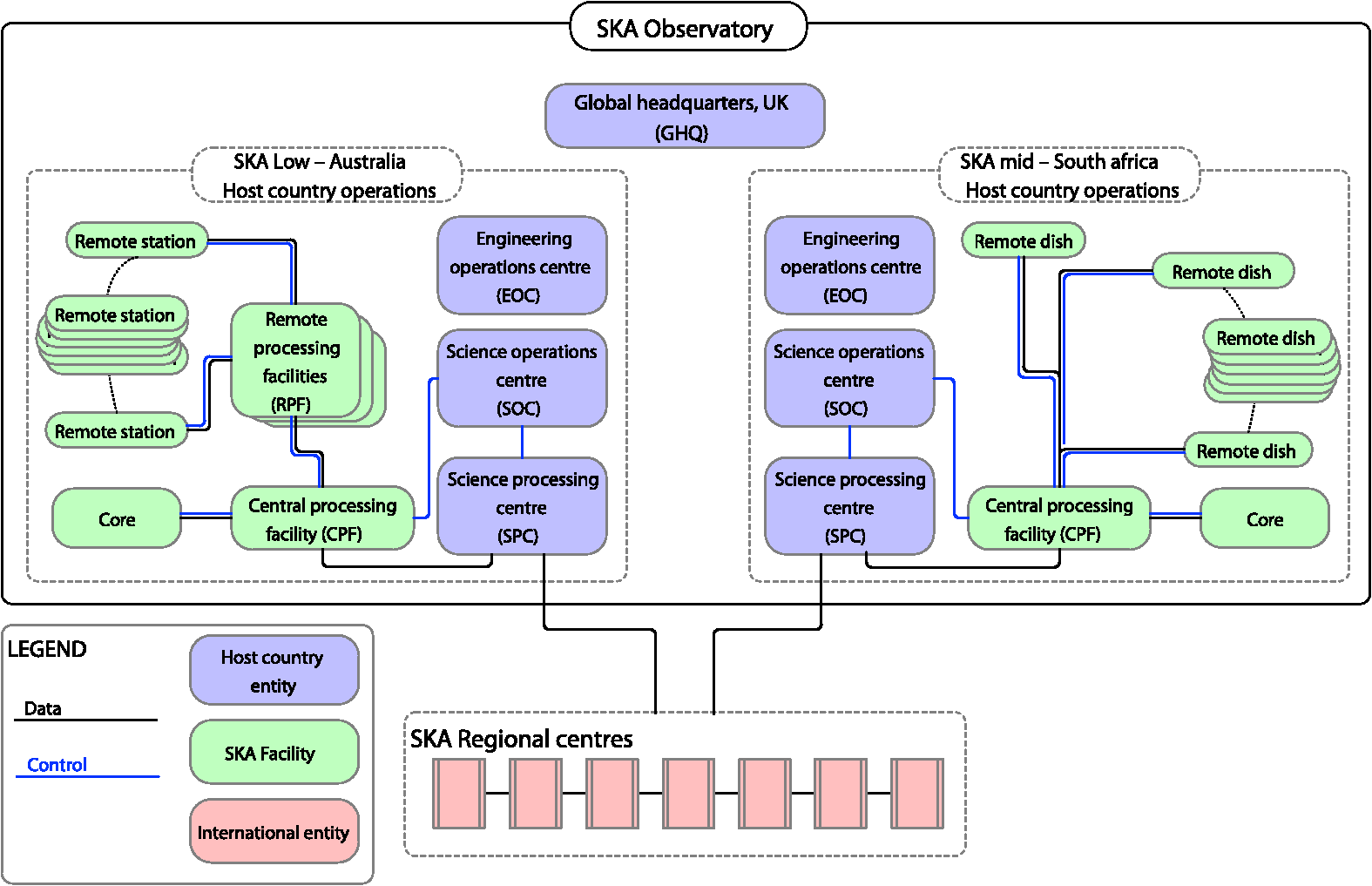}
    \caption{System of systems high level architecture of the SKAO's telescopes. SKA Regional Centres provide local access to science data products.}
    \label{fig:consortia}
\end{figure*}

The telescopes are composed of many inter-related components, each a substantial and novel engineering effort: Building the physical arrays and dishes; delivering high-speed data throughput in remote locations; securing the computing environment; maintaining data quality; and finally, managing the humans involved in building, operating, and using the system. The SKAO's telescopes are therefore complex socio-technical systems of systems. To manage this complexity, the system architecture is arranged as an ultra-large-scale system of systems \citep{FeilerUltraLargeScaleSystems2006}, depicted in Figure \ref{fig:consortia}.
Quality measures for doing science with the telescopes are similarly lofty, given the data rates and observational goals.

The SKA Project has a long timeline compared to many software and computing projects. Although the treaty that formed the SKAO came into effect in January 2021, planning and design of the telescopes started in the 1990s. The Operation Readiness Review, the project milestone when the telescopes are handed over to the operations teams, is currently planned for mid-2027. The operational lifetime of the Observatory is planned to be at least 50 years.

The software architecture designs must therefore recognize two real-world problems, namely: 
1) the telescopes will operate in a context that their initial designers could not fully understand: for example, faster processors, more efficient software, and improved networking;
2) providing this flexibility for future changes must still respect today's cost, schedule, and technical constraints, such as processing efficiencies.  

The software architecture challenges faced by SKAO, and in particular, the Science Data Processor component, are the focus of this paper. We explain the software requirements aspects of the SKAO's Science Data Processor, and explain how that software fits with the other elements (for example, the signal processing pipeline).  
We describe how we used and customized existing software architecture analysis techniques to develop the design for the Science Data Processor, and how those same techniques were also used in critical design reviews to produce risk registers representing the `known unknowns' that the project needed to address.

We begin by introducing the scientific background of radio astronomy and SKAO's telescopes needed to understand the paper\footnote{see  \url{https://www.skao.int} for more details.}.
We also introduce our architecture analysis framework, a suite of tools and methods developed at the Software Engineering Institute. 

The bulk of this paper reports our experiences applying this framework to the SKAO's telescope software design, framed as a narrative summary. We then explain the current SKA Project context and how the design approaches have held up as iterative development has proceeded. We conclude with four software engineering insights from our work on this project, and explain their implications for both researchers and practitioners.

\section{Background}
We introduce both the software architecture techniques used and the details of the organizational context to orient the reader. There is more detail here than in a typical paper but it is necessary to understand the context of this experience report. 

\subsection{The Square Kilometre Array Observatory}
\subsubsection{Science background}
Radio astronomy detects electromagnetic radiation in the centimeter- to meter-wavelength spectrum to interpret the universe. Telescope performance is measured as sensitivity and resolution. In radio astronomy performance is increased by using more receivers (grouped in an array) or longer baselines (physical separation between receivers). The scale of the SKA Project's telescopes can be appreciated in at least these four dimensions: physical size, data rate, computation, and development organization, as shown in Table \ref{tab:skao_scale}.

\begin{table*}[h]
    \centering
    \begin{tabular}{lp{10cm}}
        \toprule 
        \textbf{Dimension}
        & \textbf{Scale} \\ 
        \midrule 
        Physical size
        & SKA-Mid (South Africa)---100s of dish antennae + digital beamformer. \\ 
        & SKA-Low (Australia)---130,000 antennae grouped into 512 stations using digital beamformer. \\ 
        \midrule 
        Data Rate
        &  $>5$ terabytes per second of raw data and an estimated 300 petabytes per year of processed data \citep{Scaife2020}. \\ 
        \midrule 
        Computation
        &  Central computer for each telescope will have performance of 125 PFLOPs (assuming a computation efficiency of 10\%).\\ 
        \midrule 
        Development Organization 
        & 200 independent organizations with minimal top-down authority (discussed in the next subsection).\\
        \bottomrule
    \end{tabular} 
   \caption{SKAO Telescope Scale}
\label{tab:skao_scale}
\end{table*}

Given the development and construction costs, the telescopes are expected to operate for 50 years.

\subsubsection{Development Organization and Governance}
During the design phase, the SKAO leadership relied on influence and goodwill to coordinate the design work performed by over 200 independent organizations (largely funded by national government grants) organised into nine design consortia with each consortium responsible for the design of one system \emph{element}. Elements included the hardware components of the dishes and antennae, supercomputers for data processing, and the software to run it all. 

As noted above, these telescopes are socio-technical systems of systems, with multiple integration points and interactions among the constituents. The constituents include human systems (e.g., the teams of astronomers, academics, engineers involved in the design), software systems, and hardware systems (e.g., the physical dishes, the infrastructure to support operations such as roads and cabling).

In the scope of this experience report, SKAO's authority was primarily gate reviews for each element at the end of the design phase. Gate reviews are a leading practice on capital-intensive projects like these radio telescopes \citep{NASA:2020}. A critical design review (CDR) demonstrated that the design of an element was mature enough to proceed with physical construction and software implementation. These CDRs were high stakes events---there was no opportunity to repeat a CDR that had unsatisfactory findings.

This paper focuses on one system element, the Science Data Processor (SDP). The SDP takes relatively unprocessed data and applies algorithms (as found in, for example, CASA\footnote{\url{https://casa.nrao.edu}}) to create science data products. During our design process, we uncovered two equally important quality attribute requirements for the SDP: the software must be long-lived and maintainable, in order to support the planned 50 year lifespan of the Observatory; the software must also be able to meet some of the highest performance goals of any extant system, in order to process the data volumes this novel instrument will produce. These qualities are somewhat conflicting, so the design must make tradeoffs to strike an acceptable balance. 

\subsection{Architecture Analysis Framework}\label{scenario-framework}
Readers familiar with architecture analysis methods developed by the Software Engineering Institute (SEI) can safely skip this section.

The methods used in the engagement described in this report are based on the principle that the design of a software architecture should be shaped by quality attributes. Quality attributes are the properties of a system that stakeholders use to judge its quality. The architecture design and analysis process begins by understanding and prioritizing quality attributes, then designing structures that reflect tradeoffs among the quality attributes, and finally analyzing and evaluating the design based on the prioritized quality attributes \citep{Kazman:2012:SUS:2215086.2215263}.

By themselves, quality attribute names---scalability, performance, availability, etc.---are too broad to act on \citep{bass_software_2012}. Saying ``The telescope software should be scalable" does not tell the architect which part of the system to scale, and what level of scale is sufficient. \emph{Quality attribute scenarios} can characterize these attributes in a way that is measurable and verifiable. A quality attribute scenario consists of 6 parts:
\emph{source}: an entity that generates a stimulus;
\emph{stimulus}: a condition that affects the system;
\emph{artifact}: the part of the system that was stimulated by the stimulus;
\emph{environment}: the condition under which the stimulus occurred;
\emph{response}: the activity that results because of the stimulus; and,
\emph{response measure}: the measure by which the system's response will be evaluated.

During the engagement reported here, we used the Mission Thread Workshop, ATAM, and ad hoc discussions to elicit quality attributes as scenarios and to prioritize them, and then used the prioritized scenarios to analyze and evaluate the architecture designs. 

\subsubsection{Mission Thread Workshop}
\label{sec:mtw}
The Mission Thread Workshop (MTW) \citep{Gagliardi2013}  is a facilitated, stakeholder-centric workshop to elicit and refine end-to-end quality attribute, capability, and engineering considerations for mission threads, focusing on interoperation among system elements. A mission thread traces the data and control exchanges between elements as the processing of a stimulus event flows through the system. At each step, the architecture and engineering considerations are identified, such as interface provides/requires semantics, failure modes and error handling, resource utilization, and technical and domain-related constraints. The outputs of a MTW are quality attributes and architectural challenges, and identification of gaps in capability, functionality, documentation, and engineering.

\subsubsection{Architecture Tradeoff Analysis Method}
The Architecture Tradeoff Analysis Method\textsuperscript{\textregistered} (ATAM\textsuperscript{\textregistered}) assesses the consequences of architectural decisions in light of quality attributes and business goals~\citep{Clements:2002,Kazman:2012:SUS:2215086.2215263}. It brings together three groups: a trained evaluation team; the architecture’s decision makers (the architect, senior designers, the project manager); and representatives of the architecture’s stakeholders. Quality attributes are elicited, represented as quality attribute scenarios, and linked to business or mission goals. Scenarios are prioritized, and high priority scenarios are analyzed by having the architect explain how a system built using the architecture would satisfy the scenario. During the analysis, the evaluation team and stakeholders ask questions to identify \emph{risks} (potentially problematic architecture decisions), \emph{non-risks} (good architecture decisions), and \emph{sensitivity points} and \emph{tradeoffs} (architecture decisions that have a significant impact on one or more than one quality attribute).

\subsubsection{Views and Beyond}
Views and Beyond (V+B) is an approach to documenting a software architecture~\citep{Clements:2011}. Views are chosen to represent stakeholder concerns and used to document software structures (elements, relations, and properties of each) including interfaces and behavior. Beyond the set of documented views, architects add information to describe the overall architecture rationale and approach, and to map between views.
\begin{figure*}[tbh]
    \centering
    \includegraphics[width=.8\textwidth]{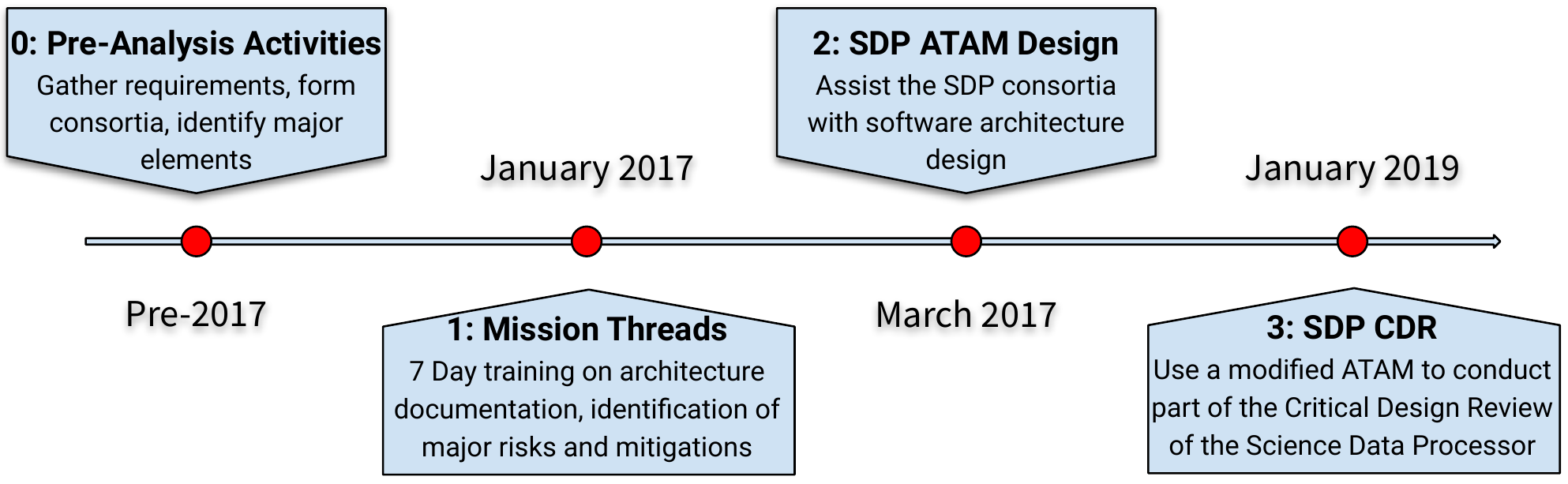}
    \caption{Steps involved in software design of the SKAO's SDP software element.}
    \label{fig:timeline}
\end{figure*}
\section{Experience Report}

This section explains how we combined the existing techniques for software design analysis and applied them to the SKAO's telescope problems. We explore how we adapted the delivery of these approaches for the unique characteristics of this system. Our efforts had four distinct phases, shown in Figure \ref{fig:timeline} and discussed in the sections that follow. Phase 0 was preparatory work. In Phase 1, we conducted system-level scenario analysis using the Mission Thread Workshop. In Phase 2, we used the ATAM to explore design alternatives for the Science Data Processor (SDP) element. Finally, in Phase  3 we used the ATAM to organize the SDP element CDR. We conclude each section reflecting on what worked well and what had to be adjusted from each technique. In Section 4, we explain the current state of the SKAO software design, and in Section 5, we derive insights for practitioners and researchers based on our experiences.

\subsection{Phase 0: Launching the Engagement}

As the overall SKAO design phase got underway, the Head of Computing and Software\footnote{At the time of the initial engagement, the first two authors (NE and JK) were researchers with the SEI. NR was Head of Computing and Software at SKAO; and MB was Software Quality Engineer at SKAO.}
was looking to reduce element integration risk. One of the few levers he controlled was specifying the  design review process and what artifacts were to be delivered. He decided to use a suite of methodologies and training developed by the Software Engineering Institute (SEI) in order to ``gain confidence (primarily through documentation) that all of these pieces of software will fit together and do what we need”. This decision was largely based on a peer recommendation%
, and the SEI's reputation. 

\subsection{Phase 1: Using the Mission Thread Workshop To Identify Challenges}\label{MTW-training}
The first engagement between the SEI experts and SKA Project Team was a Mission Thread Workshop (MTW). As noted above, a MTW is designed specifically for systems of systems integration. We used the notion of a mission thread to motivate training on architecture documentation using the Views and Beyond (V+B) discussed above---when the explanations of processing or interactions at a thread step became complicated, we identified the architecture view that would support the reasoning. The workshop and training took place over seven consecutive days in January 2017. There were more than 30 attendees representing ten countries, from all element design consortia and bringing extensive expertise in radio telescope design and operation. 

We chose three mission threads---end-to-end system control and data flows---to trace through the envisioned architecture of the telescope as a whole. For example, one thread was ``normal operation,'' describing how one of SKAO's telescopes should work: scheduling a scientist's observing time, pointing the instrument at the desired patch of sky, receiving data, processing the data, and delivering a scientific product (for example, the radio visibility values for that patch of sky over a period of time). At each step in the thread, attendees identified potential risks in the design.

In practice, many software integration issues occur at the organizational (consortia) boundaries. In SKA's case this risk was greatest between the design consortia, for example, where data is handed over from initial acquisition to the science data processor. The end-to-end scope of mission threads readily revealed these flawed assumptions.  For example, one element design incorrectly assumed that certain processing was applied before it receives a data stream. 
Although this started as a training engagement, working through SKA-specific examples and scenarios during the training proved to be an important way to motivate the approach and the artifacts produced during the training exercises delivered immediate value from the training investment.

\noindent\textbf{Outcomes/Next Steps:}
This workshop increased the SKA Project team's confidence that the plan and design might be feasible. All consortia members became familiar with SEI jargon and approach (scenarios, quality attributes, documentation templates, and so on), which would be important for the upcoming CDRs. There was a new awareness of the details of other components of the project and the project's scope, and participants made personal connections with consortia peers responsible for other components. Each consortium then went back to its own work on detailed designs and preparing for the element's CDR with SKAO. 

\subsection{Phase 2: Architecture Design and Documentation of the Science Data Processor Component}
Shortly after the initial Mission Thread Workshop with the element design teams, SEI experts met with the SDP consortium team, to provide coaching in the application of the architecture documentation and analysis methods to help the SDP team prepare for their CDR with the SKAO.

This phase began with an in-person working meeting in 
March 2017 between two SEI experts, the principal members of the SDP architecture design team, and observers from SKAO. Techniques from the ATAM (scenarios, utility tree, and prioritization voting) were used to elicit and prioritize quality attributes. The SEI team helped to document other architecture drivers, such as constraints and assumptions.

The working meeting then turned to applying the V+B architecture documentation approach. The SDP team identified architecture documentation stakeholders and documentation uses, and chose a set of views to address stakeholder needs. Stakeholders included the SDP implementers and the CDR review panel. The set of views comprised two module decomposition and dependency views (one for the science processing workflows and one for the framework that executes the workflows), two  component and connector views (data processing and security), and a data model. Several \emph{beyond views} documentation artifacts were also defined: 
a hardware decomposition and dependency view to provide context for deployment and performance analysis, a functional architecture view to provide requirements traceability, and use case views showing how science workflows are created and executed. Final versions of these architecture documents can be found at \url{https://web.archive.org/web/20221202173301/https://ska-sdp.org/publications/sdp-cdr-closeout-documentation}. 

The SDP team had already created a number of documentation artifacts, which were mapped into these chosen views. All the main artifacts that had been produced during the SDP architecture design found a place in the Views and Beyond outline.

The SDP team was having difficulty making SDP architecture concerns visible to SKAO. For example, the functional and quality attribute tradeoffs at the interface of the cluster manager and the SDP execution framework were complex and difficult to explain to non-experts. Traceability from business/mission goals to quality attributes and scenarios to architecture tradeoffs helped the SDP team expose the complexity and impact of the architecture decisions. 

\noindent\textbf{Outcomes/Next Steps:}
The SDP team reorganized their documentation artifacts into the V+B outline created at the working meeting. There were still many empty or incomplete sections in the outline, and the team then took an artifact-driven approach, performing the design and analysis to fill in the gaps in the outline.

\subsection{Phase 3: Science Data Processor pre-CDR and CDR}
\label{ATAM-at-CDR}

We then applied the ATAM to evaluate the Science Data Processor (SDP) software as part of the SDP pre-CDR. The SDP design was \textit{unprecedented} along dimensions that included data input rates, processing throughput, and data product size. SKAO wanted to allow as much time as possible for the architecture to mature, so  a pre-CDR review was held in %
June 2018 to assess the work in progress, with the final CDR deferred to January 2019.

The SKAO Head of Computing and Software led the evaluation with support from two SEI experts, one of whom was a certified ATAM Lead Evaluator who ensured that the ATAM processes were applied correctly. The evaluation team comprised four senior staff from SKAO and two external reviewers, recruited from the fields of high-performance computing and signal processing algorithm implementation. The utility tree contained more than 50 scenarios, three of which are shown in Table \ref{tab:sdp_scenarios}. The 11 highest priority scenarios were analyzed. The evaluation findings produced 57 risks, grouped into 9 risk themes. For example, one risk theme was the %
need to support data flows needed by certain types of algorithms. 

\noindent\textbf{Outcomes/Next Steps:}
The SDP architecture team worked to address the risks found in the pre-CDR evaluation. 
This involved redesign, improved documentation, and prototypes and experiments to demonstrate feasibility. SEI experts continued to provide coaching and advice to the SDP architects, and performed an active review of the nearly-final SDP architecture documentation, checking to see where each of the pre-CDR risks was addressed in the V+B architecture documents. 

The SDP team was thus well prepared to pass the final CDR, which also used the ATAM, in January 2019. The review team comprised senior SKAO staff (i.e., organizationally independent of the SDP team) and two external reviewers with combined decades of expertise in telescope construction and operation.

\begin{table*}[h]
    \centering
    \begin{tabular}{ccp{3.8cm}p{4.8cm}c} \toprule
        \textbf{Quality} & \textbf{Environment} & \textbf{Stimulus}  & \textbf{Response} & \textbf{Priority} \\ \midrule
        Modifiability & Operations & A project requires the processing of visibilities using the SDP instance on the SRC platform, using new algorithms developed by the project team. %
        & The SDP at the SRCs support generation of different instances allowing new algorithms and workflows to be combined with existing capabilities, %
        while still permitting other large-scale observing programs. & H,H \\
         Security & Operations & Unauthorized access is executed on the SDP  & Access is detected and root cause analysis information is available to understand and close problem within 1 week. & H,H\\
         Availability & Operations &  Regional Data Centre connection is degraded for a long period of time.  & Telescope continues to observe with data going to storage, minimizing impact on current schedule. Data catch-up while still operating. & H,H\\
        \bottomrule
    \end{tabular}
    \caption{Selected scenarios used in the SDP CDR. The Priority reflects mission Importance and technical Difficulty.}
    \label{tab:sdp_scenarios}
\end{table*}

\subsection{Summary of Changes to Analysis Approaches}
    Methods such as ATAM and QAW are well-defined and the product of years of engagements. 
    However, the unique context of SKAO meant we made certain adaptations.
    \begin{itemize}
        \item We integrated the reporting of ATAM risks and risk themes into the existing system engineering lifecycle at SKAO; this meant creating a list of ``Observations'' based on the documentation, which were captured as Jira tickets. A Major Observation was defined as ``anything that has a major impact on the Element requiring architectural changes or affects System Level interfaces, budgets, requirements, ...". 
        \item The ATAM findings were combined with non-software review findings to produce the overall SDP CDR recommendation. SKAO decided that the SDP passed the CDR, with a number of outstanding action items that included the ATAM risks.
        \item SKAO is a multi-national organization with no direct hierarchy. Meeting times constrained the traditional approach to architecture analysis. For example, SKAO  architecture team drafted a utility tree which was augmented by a very brief elicitation session during the meeting, rather than creating that in the meeting itself. 
        \item ATAMs and QAWs typically separate internal and external stakeholders. In the case of SKAO, this was not necessary as the stakeholders were all internal. 
        \item The Mission Thread analysis was combined with an education session in a week-long training and elicitation approach. This ensured SKAO participants were familiarized with the broader approach and scenario-based analysis.
    \end{itemize}

\section{SDP Design Maturation and Ongoing Work at SKAO}
We now expand on the design decisions taken as part of the analysis process, and describe the current state of software work at SKAO. 

\subsection{SDP Design Maturation}
\begin{figure*}
    \centering
    \includegraphics[width=\textwidth]{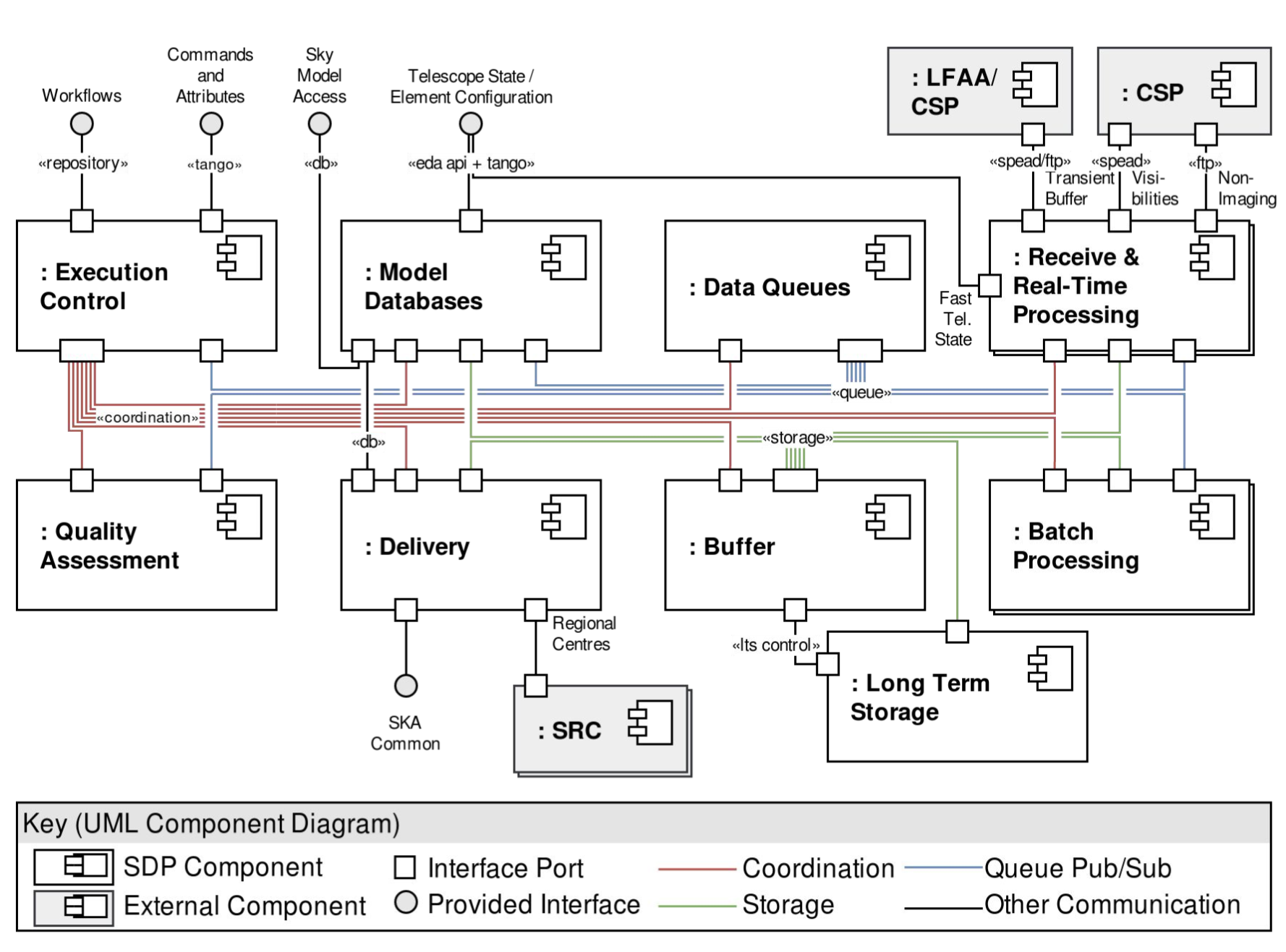}
    \caption{UML component diagram showing data flow view of the SDP architecture. The architecture tension is between the real-time processing demands of the data stream from the Central Signal Processor (upper right) and the batch processing and delivery mechanisms, all of which will evolve over the lifespan of the Observatory.}
    \label{fig:sdp-arch}
\end{figure*}

Based on the architecture analysis and quality attribute scenarios recovered, the current Science Data Processor data flow design is shown in Figure \ref{fig:sdp-arch}. \cite{Swart2022} provide a more detailed description of the entire system architecture of the MID instrument. %
The majority of data comes to a processing center hosting the SDP software from the Central Signal Processor component located near the dishes. This raw visibility information is then processed by the SDP. Other data sources include the Telescope State (e.g., which dishes are dedicated to which tasks) and models of the sky (for correlation and correction). The SDP will manage execution of multiple batch algorithms, perform quality assessment checks, and ultimately store and transfer processed data to long term storage and SKA Regional Centres (SRCs). 

A big challenge when developing a complex system with next-generation demands is to identify what aspects of the problem are \emph{precedented} and which are \emph{unprecedented}. 
For example, the telescope architects made an early decision to base their low-level system control on the Tango middleware framework \citep{pivetta2018tango}, because Tango is relatively well-known, and choosing it simplifies subsequent decision-making. 
Previous instruments and precursors have successfully used Tango, so it is precedented. Indeed, the decision to use Tango continues to be validated \cite{Swart2022}.
On the other hand, the design of the Science Data Processor element was unprecedented and wide open, and was unlike other scientific computing environments.
Implementation choices for the SDP have therefore been based on safe-to-fail experiments and prototypes to buy down risk. We now outline some of these approaches.

\subsection{Ongoing Work at SKAO}
SKAO transitioned into the construction phase in July 2021. For the software team at SKAO, the construction phase is when contracts are awarded for software development. The team has focused on preparing for collaborative development, building the tooling, testing, and continuous integration support. This includes using cloud infrastructure (based on Kubernetes) to virtualize the physical instruments to provide robust testing and monitoring insights. For example, the correlator (which synchronizes the raw signals from individual antennae) can be simulated to examine data rates and data stream characteristics. This virtualization will allow the software team to operationalize performance monitoring for handoff to hardware implementors. The team is following a process model based on the Scaled Agile Framework (SAFe)  \citep{Bartolini2020,Klaassen2020}.

The initial design documents created during the activities described in \S3 were used to successfully complete the CDR for the full system, and to create the legal entity to manage the SKAO. 
The construction plan added a new short-term milestone called AA0.5 (Array Assembly). This is a small-scale set of physical radio arrays that will allow for end-to-end proof of concept. Although the AA0.5 milestone will not provide stress testing, most of the major software components will be required. The software team used the scenarios and quality attributes described in \S 3 to determine how to adapt to this new minor increment. 

A major challenge has been analyzing distributed system failure modes, particularly infrequent anomalies (for example, those occurring fewer than 1 in 1000 events). These integration points were a focus in element CDRs, but failure modes between elements (such as between the central signal processing and science data processing) are  often  hard to characterize until implementation begins. Another challenge for the SKAO staff is managing the large number of staff and contractors responsible for different software elements, with different delivery schedules and capabilities, hosted in many different countries. 
The team is using behavior-driven testing to  align stakeholder expectations with software requirements and specification. 

\textbf{Reflecting On the Value of Architecture Reviews}
Since the CDR, several items have emerged as potential challenges that were not addressed in the initial reviews. Many of these are a result of beginning to build an executable artifact. There is a new awareness of the importance of \textbf{integration} as a key quality attribute. There is also renewed focus on reuse and use of open source software instead of implementing every solution from scratch.

A number of time-consuming concerns did not show up in design reviews. For example, the information technology aspects, such as figuring out the development stack and versioning (such as which version of Ubuntu OS). The socio-technical challenges inherent in software development are not well assessed in the ATAM-based process, which tends to examine artifacts. For example, the implementation skills and experiences of implementation teams and contractors are a big factor in technical decision making. The team may not have domain specific knowledge, or be used to working in a particular paradigm. 

There is an ongoing tension between the final, complete SKAO, and the intermediate and short-term steps to get there. The agile focus of three month release cadences from the adoption of the Scaled Agile Framework (SAFe) means diagrams are updated and redrawn as learning occurs.

Upstream, the entire SKAO has major engineering change proposals that periodically impact the software component. For example, the signal processor component originally was co-located with the antenna and dish hardware (in a remote location), and is now closer to major urban environments. Solving today's problem might possibly make the longer-term issue more difficult. The ATAM and scenario approach are not well tailored for shorter, more detailed reviews. 

Three years on from the initial approval, the SKAO architecture team sees the value in the process more than the artifacts. The ATAM and related approaches are valuable in providing a framework (the documentation style and scenario focus) that are still used as foundation for reviews and evolution to this day. There is significant value in just exposing people to a coherent architectural vision, but the vision needs to be re-emphasized to build and maintain a focus.

\section{Insights and Lessons Learned}
The observations and experiences from the CDR and design engagements led us to identify \textbf{four challenges for long-lived scientific software design}. As well as naming these challenges, we report on how  we solved each challenge on this project. We also discuss broader implications for research and practice in other, related software contexts. 
We use call-out boxes shaded in grey to summarize each of the four sections, highlighting the \textbf{problem}, \textbf{our solutions}, and \textbf{wider practice and research implications}.

\subsection{Integrating Systems Engineering and Software Engineering}
\label{sec:integrate}

Designing and integrating the massive-scale cyber-physical systems for the SKAO telescopes  requires collaboration between systems engineers and software architects. The systems engineers on this project were very experienced, acting as requirements owners and systems analysts. They decomposed system requirements and allocated them to system elements and to components within an element, and later analyzed element designs to verify that the requirements were satisfied. Requirements and interface control document (ICD) management were high priority activities during the telescope design phase to provide evidence that the proposed designs were suitable and establishing the technical contracts between consortia developing the elements.

The engagement discussed in \S 3 occurred while the telescope design team was shifting from a systems engineering perspective---treating the telescope as a collection of interacting physical elements---to a software architecture perspective focused on the interaction between the software within the elements. We observed some friction between design team members as they tried to reconcile these different perspectives. 

This challenge is not unique to this project. Systems engineering is a broad field and systems engineers may fill many roles on a project \citep{Sheard:1996}. Subsequent studies have examined the relationship between systems engineering and software architecture in the design of large-scale systems \citep{Sheard:2018,Cadavid:2020}. 

The systems engineering perspective treats the telescope as a whole, considering mechanical, electrical, software, and other types of components, and models the system as a  hierarchy with each block decomposed into sub-blocks. While this perspective serves well for physical systems, it does not easily represent software systems where services are often shared across several elements  \citep{Maier:2006}. This difference in perspectives is an essential difference between the two disciplines, and remains unresolved in practice \citep{Cadavid:2021}.

The consortium-style organization of the design teams added challenges. Teams were funded by separate national agencies and had no authority over each other. The SKAO also had limited authority, essentially only able to accept or reject designs at PDR and CDR, and the SKAO was using conformance to ICDs as a primary acceptance criterion. The element-to-element ICDs had two deficiencies: First, while the ICDs specified syntax and protocols, they did not specify behavior such as message sequence constraints or error handling. Second, each ICD focused on a single interface with no artifact representing collaboration among several elements using multiple interfaces. 

Teams did overcome the systems engineering/software friction that is frequently observed in practice \cite{Cadavid:2020}. One possible explanation is that all team members participated enthusiastically in the Mission Thread Workshop and associated training discussed above in Section \ref{sec:mtw}. 
The workshop provided a concrete bridge from systems engineering concerns to software architecture concerns and introduced a common language to discuss those concerns. The creation and elaboration of each mission thread produced an artifact that used the existing ICDs for interface details and added the collaboration among elements, both direct and mediated by other elements. The mission thread also provided the context to identify and reason about quality attribute concerns that were important to both systems engineers and software designers, such as availability, scalability, and throughput.

\noindent\textbf{Practice Implications}
The discussions during this Mission Thread Workshop were unlike any previous workshops that we have facilitated. Our experience has been that the primary functional thread is already documented or is completely understood by all participants so that any of them could quickly recite the thread steps. Typically, the workshop session focuses on quality attribute considerations at each step. In this case, the primary mission thread (i.e., how the telescope's normal operations should work) had not been documented, and the fragmented authority meant that no one was responsible to develop it. Much of the workshop time was spent developing the basic functional thread steps. As a result, we recommend deciding upfront whether the prime focus of the MTW will be to promote discussion among stakeholders or produce concrete deliverables.

\noindent\textbf{Research Implications} This project shows that further research is needed into the socio-technical interactions between systems engineers and software architects. Another research opportunity is how to move beyond ICD conformance as the design acceptance criteria. Last, trans-disciplinary research is needed into how to structure agreements and incentives in a design context like the SKA Project that has independent teams and limited global authority, and how to operationalize measures of conformance to those agreements. 
\vspace{1ex}
\begin{tcolorbox}[title=\#1 Integrating Systems Eng. and SW Eng.,boxsep=.5mm]
\textbf{Problem}: Socio-technical system of systems, with complex HW and SW components.        \\
\textbf{Solution}: Mission-thread analysis surfaces integration challenges and builds system-wide consensus.\\
\textbf{Practice}: Decide early on the principal deliverables from the workshops.\\
\textbf{Research}: Integrating systems engineering outputs like ICDs with SW eng. outputs like design documents.
\end{tcolorbox}
    
\subsection{Upskilling Domain Experts and Contextualizing Software Experts}
\label{sec:upskill}
Large-scale astronomy projects originate with astronomers, not software or systems engineers. There are complex technical challenges---such as processing of raw radio frequency signals into visibility data---that astronomers know better than anyone. At the same time, the typical astronomer has very little software engineering experience, perhaps having attended a workshop like Software Carpentry\footnote{https://software-carpentry.org}. Thus most astronomer's software expertise comes from experience, which leads to a reliance on ``how we did it before” and lessons from earlier projects. %
Furthermore, the radio astronomy community is not large, and so a large fraction of the external reviewers (for example, on a CDR  panel) share many of the same experiences with the design team being reviewed.

Similar situations are common in high-energy physics, genomics, and other science disciplines. %
Over time a science community develops highly-skilled software designers and engineers who move from project to project, but it can be hard for science projects to match the industry compensation levels needed to retain these experts.
At the same time, the domain knowledge is as hard or harder to acquire for software engineers, requiring in this case postgraduate-level knowledge of radio astronomy concepts. 

Thus, one of the challenges for the SKA Project was to simultaneously upskill radio astronomers in software architecture, and to teach non-astronomers sufficient domain knowledge to allow them to communicate effectively. This domain knowledge transfer problem exists in other industries, but the amount of specialized knowledge required is often smaller, and commercial domains like automotive now have dedicated software teams. 

Our approach to structuring the engagement was to combine existing training materials (which we had used in many previous engagements) with domain-specific examples. We formed a 3-phase approach to bridging domain experts and software experts: in the Analysis phase, the teams Identify a problem in their existing system/process; in a Training phase, the consultants then help participants Learn a method to fix that problem; and finally, in the Practice phase, participants collaboratively fix the problem, with coaching from the instructors.

\noindent\textbf{Practice Implications}: 
One benefit of recruiting generalists to help with architecture analysis or review is that an ``informed ignoramus'' can question assumptions or simplify the problem. However, there are limits to what a generalist can contribute. On this project, we (generalists from the SEI) saw cases where the review scope might exclude what we thought were important questions. For example, the need for high levels of resiliency or for rapid failure recovery are less important for this system, since most observations can be restarted and repeated. 

Thus, on this project it was hard to get designers to pay attention to failure and recovery, including concepts such as cascading failures and looking beyond ``normal operation” in scenarios. These are not interesting to domain experts, who are focused on normal operations and whether the instrument will resolve the important science questions, and less interested in, for example, securing the supercomputer from malware planted by cryptocurrency miners.

In its current phase, the SKAO software and engineering  teams have realized the value of socializing a system-wide perspective across the entire group. Workshops on use cases and technical leadership have helped build a stronger organizational culture that understands the discipline of software engineering in the context of the wider system.

\noindent\textbf{Research Implications}: There is a perennial challenge of inculcating software expertise and awareness in other disciplines. More importantly, the challenge is also how to bring existing approaches, such as requirements elicitation techniques, into consulting and collaboration exercises. Many of today's upskilling initiatives focus on the most basic software engineering concepts such as version control and testing, but other software concerns, such as release engineering, software design, and requirements engineering are as important in the early design of large-scale systems.

\vspace{1ex}         
\begin{tcolorbox}[title=\#2 Upskilling,boxsep=.5mm]
\textbf{Problem}: Designers are domain experts with limited formal SW eng. training.       \\
\textbf{Solution}: Mission-thread analysis surfaces integration challenges and builds system-wide consensus.\\
\textbf{Practice}: Need outside experts who may have limited domain knowledge. Fast \& adaptive training program to normalize terminology \& deliverables.\\
\textbf{Research}: How to upskill in areas beyond coding, version control, e.g., to include requirements and architecture.   
\end{tcolorbox}

\subsection{Risk Reduction in the Design of Long-Lived Software}
\label{sec:risk}
The SKAO's telescopes have a long temporal scale: a planned lifespan of 50 years and a major upgrade anticipated before the end of that time. This is far longer than the planned design lifespan of most commercial software systems.\footnote{although some may well persist, if unintentionally, that long, for example, COBOL-based government systems}. %
For example, Google has developed three different file system architectures since its founding 23 years ago. 

In addition to temporal scale, SKAO's telescopes have significant scale physically: SKAO is planned to consume more processing resources (100s of petaFLOPS), bandwidth (8 Tbit/s), and storage (600 Pb) than any current system, in order to process raw radio signals into science data products. 

A key activity in designing for these scales is \emph{risk reduction}: Identify the potential risks to project success, and find suitable mitigations for those risks. On this project, past experience and knowledge of radio astronomy make it easier to buy down hardware and physics based risks: The SKAO's telescopes build on the knowledge from previous telescopes such as ALMA and pathfinding prototypes such as ASKAP\footnote{Australia SKA Pathfinder, \url{https://www.atnf.csiro.au/projects/askap/}}, and on a long history of radio physics knowledge such as Fourier transforms, deconvolution, and calibration. 

Risks due to scale are harder to mitigate. 
Testing the functional correctness of scientific software is very difficult. These systems must detect small and ephemeral signals, with confidence levels of 5 standard deviations in the hypothesis tests being common in some domains \footnote{\url{https://blogs.scientificamerican.com/observations/five-sigmawhats-that/}}. The telescope systems on this project must also contend with testing at scale: Software works differently when integrating 100,000 instances of a component working together, compared to integrating 10 instances. Finally, at this scale, low probability and intermittent hardware and software failures become everyday occurences.

It is also conceptually easier to model and identify risks associated with hardware components. 
Software is often seen, especially by non-software engineers, as infinitely malleable, weight-free and cost-free. Engineers make implicit assumptions about the evolvability of hardware vs. software (i.e., that Moore’s law means exponential increase in processing speed). 
Physics risk can also be easier to quantify; we can mathematically model antenna pointing error bounds using the fabrication and assembly tolerances of the physical components. 
In software, by contrast, such models are significantly harder to construct and analyze. We have good tools for proving algorithmic properties (for example, the running time of a processing algorithm), but these tend to cover a small subset of the overall software system. 
For example, we can make only rough estimates to quantify the risks of unforeseen bugs, amount of development effort, or the longevity of key libraries. 
Using a quality attribute scenario approach in ATAMs and Mission Thread Workshops supports this risk-focused approach; the goal of a ATAM is to identify risk themes in the current design approach.

Risks can also arise from the design process itself. This project, for example, is currently using an agile approach with the Scaled Agile Framework (SAFe) \citep{Bartolini2020}, however this approach is unproven at SKA Project's scale. These software systems are being designed 10-15 years before production deployment; this means accurate, at-scale feedback will not be available for years, and key design decisions may not be tested in production for decades. Software architecture embodies design decisions that are expensive to change~\citep{Booch:2008}; it is possible to rewrite code, but reverting key design choices can be nearly impossible. 

The SKA Project design teams must live with this risk: there are still unknown unknowns that cannot be addressed or identified, and the only approach to take is to experiment, learn lessons, and adapt.

\noindent\textbf{Practice Implications}: The SKA Project uses four key approaches for risk reduction at scale. First, the roll-out plan calls for the telescopes to be incrementally delivered in four Array Assemblies, with increasing numbers of receptors and validated capabilities; second, development will use a modified incremental and iterative approach to learn lessons by continually delivering some form of working software, even if the software does not go into production; third, designs use open source software where possible, to avoid lock-in and vendor restrictions (for example, the use of Tango middleware); finally, the project is prepared to take ownership when needed, in case the open source project maintainer quits or retires.

\noindent\textbf{Research Implications}: There are several research implications of our risk reduction approach. One is how well  large-scale agile, such as SAFe, works compared with earlier methodologies like the Spiral model or iterative and incremental development. Is large-scale agile simply a misnomer? How do you apply agile principles to long-lived projects? 

Software engineering has long struggled with how to build knowledge and confidence in a design approach before beginning implementation. Currently this is resolved by shortening the cycle time from design to production as much as possible, but in many projects, like SKA, cycle times are necessarily long. Tests in production do not work if the user base does not yet exist, and using synthetic data is not adequate, particularly in scientific data processing. 

Model-driven approaches, most recently in the form of digital twins, promise to make such large-scale simulations possible. Facebook, for example, has a digital twin of its entire social graph in order to test new optimizations for its product offerings \citep{Ahlgren2021}. Digital twins are also proposed for mechatronic systems that combine hardware and software components.

\begin{tcolorbox}[title=\#3 Risk Reduction for Long-Lived Systems,boxsep=.5mm]
\textbf{Problem}: 50-year planned lifetime, 10-20 year design timeline. Risk management easier for HW components.       \\
\textbf{Solution}: Use scenarios in ATAM to focus on risk themes and create risk registers.        \\
\textbf{Practice}: Use an iterative and adaptive approach focusing on standards and open source software.\\
\textbf{Research}: Track \& analyze success of large-scale iterative models. Virtualization \& twinning may reduce risk of longer increments.   
\end{tcolorbox}

\subsection{Use of Scenario-based Methods}
\label{sec:scenario}
Many of the architects on this project played the stakeholder role of designer \textit{and} the role of domain expert \textit{and} the role of stakeholder as eventual user or operator of the system. This led to the need to represent and distinguish diverse types of concerns, for example, technology tradeoffs, physics limitations, and quality attributes. In this large and distributed project, clearly communicating such concerns across design teams was a challenge.

While software architects spend a significant portion of their time communicating \citep{Kruchten:2008}, the relative inexperience of the architects and the diversity of the audience made the situation here challenging. For example, as discussed earlier, the SDP team had difficulty making core tradeoffs (such as at interface boundaries) visible to reviewers. 

Scenarios, described in \S \ref{scenario-framework}, became an important tool for communication. The template leads an inexperienced architect to clearly frame an architectural concern, showing the need to make a decision or tradeoff, and providing a yardstick to measure each alternative. Each scenario was mapped to a business or mission goal, which defined the significance of the concern. The broad training discussed in \S \ref{MTW-training} ensured that other designers would understand the framing of concerns as scenarios. 

In the SDP example, the team created scenarios to illustrate the impact of the tradeoffs, mapped each scenario to operational and science goals, and then used the scenarios to frame the tradeoffs to SKAO and reach consensus on  technology decisions.

The scenario-based MTW and ATAM methods embody fundamental steps: Identify business/mission goals, elicit scenarios and map each to goals, prioritize and analyze the ability of the architecture to satisfy high-priority scenarios, identify risks, and feedback findings and iterate the steps. These steps were performed to communicate the architecture, as described in this section, and to evaluate the architecture, as described in \S \ref{ATAM-at-CDR}. 

Although scenarios were a valuable tool to improve communication, the inexperienced architects still had to triage, prioritize, and make innumerable decisions. 

\noindent\textbf{Practice Implications}: Scenarios embody a common, accessible narrative that resonates with many stakeholders. However, prioritizing and triaging these scenarios is largely qualitative and highly dependent on whom is surveyed.

\noindent\textbf{Research Implications}: Methods are needed to guide less experienced architects, in particular guidance on how to sequence design decisions. This project's context is particularly stressing, with several years between separate design and construction phases, creating a long feedback cycle of final evidence of design quality. Principles are needed to guide the breadth and depth of the analysis and evaluation activities, to make the process less dependent on the expertise of the analysts.
      
\begin{tcolorbox}[title=\#4 Scenarios,boxsep=.5mm]
\textbf{Problem}: Inexperienced architects with diverse concerns.  \\
\textbf{Solution}: Drive communication via scenarios.\\
\textbf{Practice}: Prioritization still a challenge. \\
\textbf{Research}: Sequencing design decisions; shorten design quality feedback; reduce process dependence on experts.   
\end{tcolorbox}

\section{Limitations}
As a singleton case study, the work described here is specific to the context and circumstances of the SKAO project, as well as the expertise of the people involved. However, the general approach of applying scenarios to architecture design and analysis has been established in other settings. We feel the specific details of radio astronomy have several lessons for other software systems that feature high data volumes, long life spans, and complex organizational structures. 
Similarly, while we know the SEI approaches best (MTW, ATAM, etc.) other risk-driven, scenario-based approaches should work as well, such as the Rational Unified Process \citep{kruchten2004rational}. None have the lengthy track record of the ones we describe here, however. 

Since the SKAO's telescopes are not yet operational, the designs and approach are speculative until realized; we do not claim that the software design is perfect but rather that the design is a reasonable one, and that any risks are acceptable. And, naturally, the design can change as unanticipated factors arise.

\section{Related Work}
\noindent\textbf{Design and Architecture Approaches for Software Systems} have been extensively researched and numerous techniques, tested in practical settings, exist. A good survey of the human aspects is \cite{Tang_2017}.\cite{Maranzano2005ArchitectureReviewsPractice} survey the practice of architecture reviews. For a perspective on system of systems approaches, \cite{Klein2013} is a good place to begin. The ATAM and related work mentioned in this paper are articulated and combined in the book \textit{Software Architecture in Practice} \citep{bass_software_2012}.  \cite{Bellomo2015TowardAgileArchitecture} summarize the experiences of applying ATAM across several decades and hundreds of projects, finding  \emph{maintainability} to be a key concern, as was the case for SKAO. Several experience reports describe how large-scale systems are analyzed, such as \cite{Rueckert2019}, which looked at decision forces on the architecture of a large industrial control system, and \cite{Bucaioni_2021}, explaining how to align business goals and quality attributes for over the air updates of automotive software. Closely related to this study is the report from \cite{Cadavid2022SystemSoftwareArchitecting}, which explores system of systems issues in a different component of the SKA telescopes. Like us, they identify integration issues as a major challenge.

\noindent\textbf{Design In Scientific Software}:
For a more general background on software engineering for science we refer the reader to \cite{Heaton-Carver:2015} and ~\cite{Wilson2014}, and the workshop series SE for Science.\footnote{\url{https://se4science.org/workshops/}}

Most scientific computing, for example, in epidemiology, energy system models, or biology, often has important software components but relatively little devoted architectural effort behind these components. A significant amount of scientific software is focused on modeling and simulation, for example, of the Earth's climate. SKAO, by contrast, will be primarily a data collection and analysis platform, with results that may feed into scientific models (e.g., of how pulsars are formed). These modeling-oriented projects tend to evolve organically as models are themselves evolved, as described in \cite{Easterbrook2009}, as opposed to dedicated instrument software, as in SKAO, which is typically more structured.

These efforts tend to lack dedicated and professionally-trained software engineers who are engaged on a long-term (rather than annual contract) basis. Recently the term Research Software Engineer has been coined to describe this role \citep{Prause_2010}.

Several communities in science have pioneered disciplined \emph{engineering} of software systems. 
CERN, the EU collaboration for nuclear research, has run a series of experiments based on world-leading software systems for decades (for example, \citep{azzopardi2019software}). Easterbrook and Johns reported on the approach taken in climatology \citep{Easterbrook2009} to build, maintain, and update the Hadley climate model code. On these projects, dedicated software engineers work closely with climate scientists to write the software, typically implemented in Fortran, to make the models efficient and scientifically accurate.

In astronomy, the need to manage observation schedules, process large volumes of data, and distribute and analyze that data, has led to software of high quality with maintenance cycles that rival large industry leaders like Google or Facebook.
The most relevant work is experience reports from other telescopes---The community is fairly small and lessons learned are widely shared. 
For example, the Atacama Large Millimetre Array (ALMA), engaged in similar science to the SKA Project, captured its experiences in reports describing how the software was created, challenges faced, and the approaches taken \citep{Chavan2012,Marson2016}. 
In addition, the ALMA project hosts open source software for managing the telescope scheduling, and many of those lessons (both good and bad) informed the SKAO's telescope designs, in some cases by having ALMA experts advise the project. There are, however, relatively few reports on applying architecture tradeoff analysis in the systematic way described here. In the systems engineering domain, \cite{Cadavid:2021} conducted a series of interviews with the Low Frequency Array (LOFAR) radio telescope architects to examine the gap between software and systems engineers. 

Many of the SKA Project's challenges have been documented as part of the project's open development process (available at \url{skatelescope.org}) or in experience reports published in venues of the astronomy community. For example, \cite{Baffa:2019aa} describe the telescope monitoring infrastructure experiments, and \cite{Bridger:2017aa} report on the task scheduling design approach and investigation. In software engineering terminology, these experience reports and prototype experiments on telescope design elements are architecture spikes or tracer bullets \citep{Hunt:1999zp} that uncover unknowns and buy down risk.

\section{Conclusion}
The SKA Observatory is a unique, high-reward endeavor to see farther than humans have before. Building the Observatory is a socio-technical challenge that requires managing complex system of systems problems, not least of which is the unprecedented scale of the data ingest, processing, and sharing. Hundreds of highly skilled people are involved. Applying tested architectural analysis approaches has been able to reduce the known risks and uncover some of the unknown risks, and prepare the software aspects of the Observatory for the construction and operations phases. In doing so, we reported on several interesting challenges, such as how to upskill domain experts in software engineering, and software engineering over very long time scales.

\section{Author Statement}
\textbf{Neil Ernst} 
Data curation,
Formal Analysis,
Investigation,
Methodology,
Software,
Validation,
Visualization,
Writing – original draft,
Writing – review \& editing.
\textbf{John Klein}
Conceptualization,
Data curation,
Formal Analysis,
Funding acquisition,
Investigation,
Methodology,
Project administration,
Resources,
Software,
Supervision,
Writing – original draft,
Writing – review \& editing.
\textbf{Marco Bartolini}
Data curation,
Formal Analysis,
Investigation,
Project administration,
Resources,
Validation,
Visualization,
Writing – original draft,
Writing – review \& editing.
\textbf{Nick Rees}
Resources,
Conceptualization,
Funding acquisition,
Validation,
Writing – review \& editing,
Supervision.
\textbf{Jeremy Coles}
Writing – review \& editing,
Validation,
Funding acquisition.

\section*{Data Availability Statement}
The processed data informing some of the above findings are available to download from doi:10.5281/zenodo.7868987. The final documentation of the SDP CDR are available at \url{https://web.archive.org/web/20221202173301/https://ska-sdp.org/publications/sdp-cdr-closeout-documentation}. 

\section{Acknowledgements}
The authors thank the University of Cambridge and STFC for funding the SEI involvement in the SDP and the principals of the SDP design consortia---Paul Alexander in particular---for their continued support. 
Thanks also to all the participants in the workshops for their time and energy, and to Mary Popeck for her help in the second workshop. NR would also like to thank Peter Braam of Braam Research LLC for suggesting the SEI approach to him in the first place.

Copyright 2022 Carnegie Mellon University, Marco Bartolini, Neil Ernst, and Nick Rees.
References herein to any specific commercial product, process, or service by trade name, trade mark, manufacturer, or otherwise, does not necessarily constitute or imply its endorsement, recommendation, or favoring by Carnegie Mellon University or its Software Engineering Institute.

NO WARRANTY. THIS CARNEGIE MELLON UNIVERSITY AND SOFTWARE ENGINEERING INSTITUTE MATERIAL IS FURNISHED ON AN "AS-IS" BASIS. CARNEGIE MELLON UNIVERSITY MAKES NO WARRANTIES OF ANY KIND, EITHER EXPRESSED OR IMPLIED, AS TO ANY MATTER INCLUDING, BUT NOT LIMITED TO, WARRANTY OF FITNESS FOR PURPOSE OR MERCHANTABILITY, EXCLUSIVITY, OR RESULTS OBTAINED FROM USE OF THE MATERIAL. CARNEGIE MELLON UNIVERSITY DOES NOT MAKE ANY WARRANTY OF ANY KIND WITH RESPECT TO FREEDOM FROM PATENT, TRADEMARK, OR COPYRIGHT INFRINGEMENT.

Architecture Tradeoff Analysis Method® is registered in the U.S. Patent and Trademark Office by Carnegie Mellon University.

DM22-0316

\bibliographystyle{elsarticle-harv}
\bibliography{ska}

\end{document}